\documentclass[useAMS,usenatbib]{mn2e}
\usepackage{graphicx,amssymb}

\title[XMM-$Newton$ EPIC and OM observation of Nova Centauri 1986 (V842 Cen).]{XMM-$Newton$ EPIC and OM observation of Nova Centauri 1986 (V842 Cen).}
\author[Luna et al.]{G. J. M. Luna$^{1,2}$, M. P. Diaz$^{3}$, N. S. Brickhouse$^{2}$ and M. Moraes$^{3}$\\
$^{1}$ Instituto de Astronom\'ia y F\'isica del Espacio (CONICET-UBA), Casilla de Correo 67 - Suc. 28 (C1428ZAA)  CABA - Argentina.\\
$^{2}$ Harvard-Smithsonian Center for Astrophysics, 60 Garden St., Cambridge, MA, 02138.\\
$^{3}$ IAG, Universidade de S\~ao Paulo, Rua do Mat\~ao, 1226, S\~ao Paulo, SP 05508-900, Brazil. }
\begin{document}

\date{Accepted 2012 March 19. Received 2012 January 19}

\pagerange{\pageref{firstpage}--\pageref{lastpage}} \pubyear{20xx}

\maketitle

\label{firstpage}

\begin{abstract}
We report the results from the temporal and spectral analysis of an
XMM-$Newton$ observation of Nova Centauri 1986 (V842~Cen). We detect a
period at 3.51$\pm$0.4 h in the EPIC data and at 4.0$\pm$0.8 h in the
OM data. The X-ray spectrum is consistent with the emission from an
absorbed thin thermal plasma with a temperature distribution given by
an isobaric cooling flow. The maximum temperature of the cooling flow
model is $kT_{max}=43_{-12}^{+23}$ keV. Such a high temperature can be
reached in a shocked region and, given the periodicity detected, most
likely arises in a magnetically-channelled accretion flow
characteristic of intermediate polars.  The pulsed fraction of the
3.51 h modulation decreases with energy as observed in the X-ray light
curves of magnetic CVs, possibly due either to occultation of the
accretion column by the white dwarf body or phase--dependent to
absorption. We do not find the 57 s white dwarf spin period, with a pulse amplitude of 4 mmag, reported
by \citet{woudt} either in the Optical Monitor (OM) data, which are
sensitive to pulse amplitudes $\gtrsim$ 0.03 magnitudes, or the EPIC
data, sensitive to pulse fractions $p \gtrsim$14$\pm$2\%.





\end{abstract}


\section[]{Introduction}


Cataclysmic variables (CVs) are binary systems consisting of a white
dwarf primary and a secondary that transfers mass to it through an
accretion disk around the white dwarf. Their brightness can change by
large factors (several millions) during nova outbursts, which are due
to thermonuclear runaways of the hydrogen-rich material that has
accreted onto the white dwarf.

The power spectra of CVs display several different periodicities
which are attributed to the orbital period, white dwarf spin period
(in the case of magnetic CVs) and superhumps (which are photometric
modulations near, but not at, the orbital period) among others.
Among magnetic CVs, intermediate polars (IPs) are moderately strong
X-ray sources ($L_{X} \sim$10$^{31-33}$ erg s$^{-1}$), in which
accreting material is magnetically channeled and shock-heated in the
accretion column, producing high energy radiation as it cools before
reaching the white dwarf surface. Strong X-ray modulation is observed
at the spin period, likely due to self-occultation of the accretion
column by the white dwarf's body and/or phase-dependent absorption by
pre-shock material \citep[e.g.,][]{alan}.

Nova Centauri 1986 (V842~Cen) underwent its latest nova outburst in
February 1986 and took 48 days to decline its brightness by 3
magnitudes, which suggests that it was a moderately fast nova
\citep{sekiguchi89}. 
\citet{downes00} imaged the nova ejecta in 1998 using
H$_{\beta}$ and [O III]$\lambda$5007 narrowband filters.
The ejecta show a fairly spherical shape in both
filters (5.6$^{\prime\prime} \times$ 6.0$^{\prime\prime}$). Given this
information, therefore, nothing is particularly peculiar about this
nova. However, optical photometry observations in 2008 overturned this
picture when a fast pulsation of 57 seconds was discovered by
\citet{woudt} and attributed to the spin period of the white dwarf,
which would make V842~Cen a member of the IP class. This finding has
strong implications:
its white dwarf would be the fastest rotator currently known in a nova
remnant; and, its symmetric shell morphology would  now present a puzzle, since a
magnetic field anchored to a rapidly rotating white dwarf would be
likely to produce asymmetric ejecta shapes during a nova event \citep{fiedler80,livio95}.

In this letter we show the results of an XMM-$Newton$ X-ray and
optical observation aimed to search for the 57~s spin period
discovered by \citet{woudt} and thus confirm the IP nature of
V842~Cen. 
In Section \ref{sec:data}
we present the details of the observation and data processing, while in Section
\ref{sec:results} we show the results obtained from the timing and
spectral analysis. Discussion and
conclusions are presented in Section \ref{sec:disc}.

\section[]{Observations and analysis}
\label{sec:data}

We observed V842~Cen with the XMM-$Newton$ {\it Observatory} on 2011 February
24 for 56.9 ks using the EPIC instrument, operated in full
window mode with the medium thickness filter and the Optical Monitor
(OM) in fast mode. After removing events at periods with high flaring
particle background using a 3$\sigma$ clipping method, the resulting
exposure time reduced to 49.8 ks.

The source spectra and light curves were accumulated from circular
regions of 32$^{\prime\prime}$ and 20$^{\prime\prime}$, respectively,
centered on V842~Cen ($\alpha$=14h 35m 52.55s, $\delta$=-57$^{\circ}$
37$^{\prime}$ 35.3$^{\prime\prime}$). The background spectra and light
curves were extracted from a source-free region on the same chip taken
within a circle of 40$^{\prime\prime}$ radius. 
The total number of counts in the source region in the pn, MOS1 and MOS2 cameras was 5354 which includes 735 counts from background. We used the
$\texttt{rmfgen}$ and $\texttt{arfgen}$ to build the redistribution
matrices and ancillary responses, respectively. The resulting X-ray spectra were grouped with a minimum of 15 counts per energy bin. 
For timing analysis,
we converted photon arrival times to the solar system barycenter using
the SAS task $\texttt{barycen}$. The exposure times in the OM filters were 16594, 16599 and 16596 seconds for the V, U and B respectively.
The OM light curves were
reprocessed using the script \texttt{omfchain} with time bin size of
0.5 s.

\section[]{Results}
\label{sec:results}

\subsection{Timing}
\label{sec:timing}



\begin{figure}
 \includegraphics[scale=0.48]{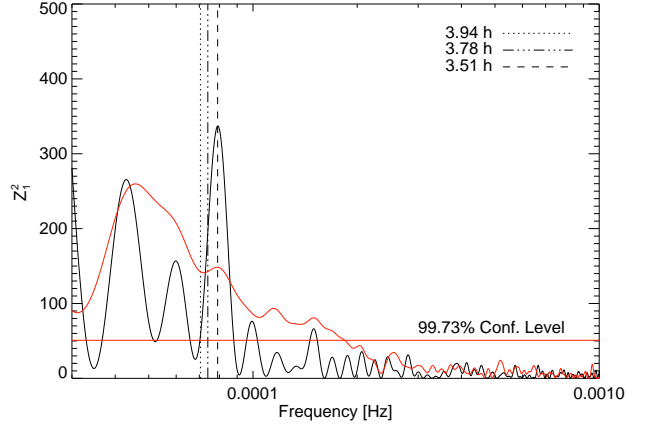}
\caption{Power spectrum $Z_{1}^{2}(f)$ from the source+background (black) and background (red) photon arrival times from the $pn$+MOS1,2 cameras. The 99.73\% (3$\sigma$) confidence level is marked by the horizontal line (red). Vertical dotted, dash and dash-dotted lines show the 3.94, 3.78 and 3.51 h respectively. {\it A color version of this figure is available online.}}
 \label{fig:pow}
\end{figure}

For the X-ray data, we used the event arrival times and
Rayleigh statistics \citep[$Z_{1}^{2}$;][]{buccheri,stute11} to
search for periods in the frequency range from $f_{min}=1/T_{span}$ to
$f_{max}=1/2t_{frame}$ with $\Delta f=1/AT_{span}$, where $T_{span}$
is the total on-source time of 49.8 ks, $t_{frame}$ is the EPIC camera
readout time (73 ms) and $A$ is the oversampling factor, which we set
equal to 1,000.  We detect a period at 3.64$\pm$0.46 h which is
within the errors of the
$\sim$3.94 h orbital period (Fig. \ref{fig:pow})  estimated by
\citet{woudt}. We are not able to detect the $\sim$57 s white dwarf spin period in the X-ray
data.

\begin{figure}
\includegraphics[scale=1]{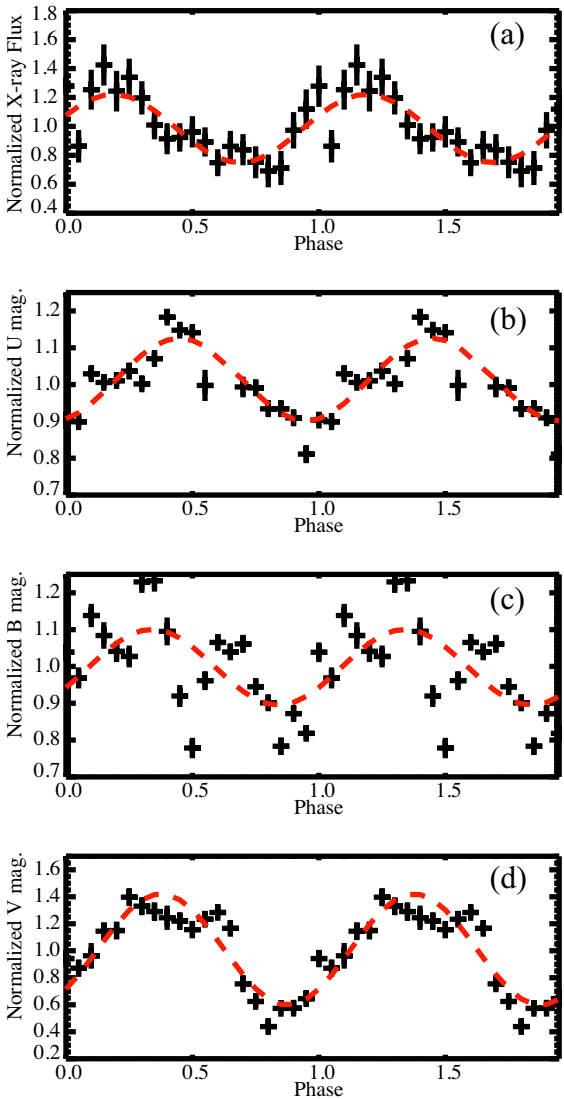}
\caption{({\it a}) EPIC/$pn$ light curve in the 0.3-10. keV energy range folded at the 3.51 h period. Normalized U({\it b}), B({\it c}) and V({\it d}) magnitude light curves folded at the 3.51 h period. A sine wave fit of the form $A+B\times sin[2\pi(\phi_{3.51}-\phi_{0})]$, where $\phi_{3.51}$ is the phase at the 3.51 h period, is overplotted ({\it dashed} line). Each point corresponds to an average over a 0.05 phase bin. {\it A color version of this figure is available online.}}
 \label{fig:lc}
\end{figure}

Since the particle background is not expected to be modulated, it
should be safe to use all the exposure time 
(56.9 ks) to improve
the accuracy of the period detection. In this case, the modulation is
found to have a period of 3.51$\pm$0.40 h.  These data are
sensitive to pulse fractions\footnote{We define
the pulsed fraction $p$ as $max/min=(A+B)/(A-B)$, where $A$ and $B$ are
determined from a sine wave fit of the form $A+B\times
sin[2\pi(\phi_{3.51}-\phi_{0})]$, where $\phi_{3.51}$ is the phase at
the 3.51 h period.} $p$=14$\pm$2\% \citep[using eq. 3
from][]{stute11}. In Figure \ref{fig:lc}a we show the $pn$ light curve
in the 0.3-10.0 keV energy range phased at the 3.51 h period. We
choose the ephemeris T$_{0}$=HJD 2454514.54882, which is the date of
the first photometric observation reported by \cite{woudt}. Due to the period uncertainty, our phase scale is arbitrary and cannot be compared with previous works.


The U, B and V light curves were scanned for periodicities ranging
from 1 s to 5 hours. Fourier transform and phase dispersion
minimization algorithms were used in order to detect sinusoidal as well
as complex pulse profiles.  No pulsed signal with semi-amplitude
above 0.03 mag could be found in the 1 to 1000s period range in U, B
and V bands. In particular, no significant peak appears at the 57 s 
period attributed to the white dwarf spin; however, we note that the pulse amplitude of
4 mmag measured by \citep{woudt} is below our detection
limits. The most significant structures in the mid-frequency range
are located in the range between 27 and 32 minutes. Their coherence could not be
verified and they may just represent the flickering time-scale in this
system. After rebinning to 100 s integrations, a 4.0$\pm$0.8 h period
is clearly seen in our light curves, with one complete cycle covered
in U, B and V on different occasions. Simulations show 
that a sinusoidal semi-amplitude of
0.13, 0.10 and 0.14 mag can be measured in U, B and V bands, respectively, for a
4.0 h hour trial period. Phase-folded light curves in the OM filters
shows significant departures from a sinusoidal shape that may be due to
flickering activity during the cycle sampled by our observations
(Figure \ref{fig:lc}). The light curves in each filter show significant change between cycles at particular phases. These changes appear as regions of larger scatter in the phase-binned curves.

The individual OM light curves sample less than two photometric cycles. Therefore, the broad peak at low frequencies (with maximum between 3 and 4 hours) includes the 2.889 h period claimed by \cite{woudt} at a high power level, producing reasonable phase-folded light curves. However, longer optical datasets are needed to study the orbital/superhump modulations and the low frequency domain.

\subsection{Spectrum}

\begin{figure}
\includegraphics[scale=0.32, angle=-90]{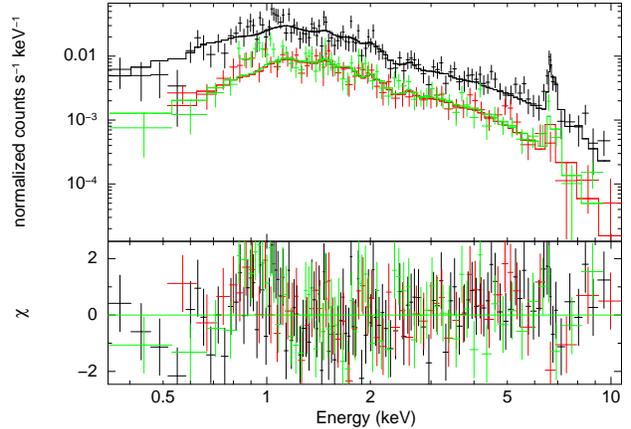}
\caption{{\it XMM} EPIC background-subtracted spectra ($pn$ in black,
  MOS 1 in red and MOS 2 in green).  The best-fit absorbed cooling
  flow model is overplotted in the top panel. Bottom panel shows fit
  residuals. {\it A color version of this figure is available online.}}
 \label{fig:esp}
\end{figure}

The X-ray spectrum of V842~Cen is relatively hard with photon
energies up to $\gtrsim$ 10 keV (Figure \ref{fig:esp}). Visual
inspection shows an  emission feature 
at energies $\sim$ 6.7-7.0 keV
which corresponds to Fe XXV and Fe XXVI. 
A simple thermal model ($\chi^{2}$=297, 271 d.o.f.) of an absorbed plasma
($\texttt{wabs}$ $\times$ $\texttt{APEC}$ in
XSPEC\footnote{http://heasarc.gsfc.nasa.gov/docs/xanadu/xspec/; $\texttt{APEC}$
models from \citet{smith01}; $\texttt{wabs}$ model from 
\citet{mmc83}}) yields
a temperature $kT=8.1_{-1.0}^{+1.6}$ keV, an absorption column density
$n_{H}=0.25_{-0.03}^{+0.03}\times$10$^{22}$ cm$^{-2}$, and abundance
$A=2.0_{-0.5}^{+0.7}$ in solar units.


Given the detection of a modulation in the X-ray light curves, we also fit the spectrum with models successfully applied 
to cooling accretion shock columns, in which the
temperature distribution depends on the cooling mechanisms \citep[e.g.,][]{alan}. In the
simplest scenario, the X-ray spectrum can be modeled with an isobaric
cooling flow \citep[e.g., ][]{mukai03,pandel05}. We use the
\texttt{mkcflow} model (with the $\texttt{switch}$ parameter equal 2,
meaning that the model spectrum used the
$\texttt{APEC}$ emissivities) modified by a simple
absorber (\texttt{wabs}). The model fit ($\chi^{2}$=260, 267 d.o.f.) results
in a maximum plasma temperature $kT_{max}=43_{-12}^{+23}$ keV,
a minimum temperature $kT_{min} \lesssim 0.6$ keV, absorption column
$n_{H}=0.30_{-0.04}^{+0.04}\times$10$^{22}$ cm$^{-2}$, abundance $A
\geq 2$ in solar units and mass accretion rate
$\dot{M}$[10$^{-12}$]=7$_{-2}^{+2} M_{\odot}$/yr ($d$/1.3 kpc)$^{2}$
\citep[using the distance from][] 
{gill98}.

At the accretion shock front,
the shock temperature is proportional to the
square of the free-fall velocity, which in turn is set by the mass of the white
dwarf. Thus the white dwarf mass can be derived from the value of $kT_{max}$
obtained from fitting the spectrum with a cooling flow model
\citep[e.g.,][]{yuasa}. 
The value of
$kT_{max}$ obtained from our fit implies $M_{WD}$=0.88 M$_{\odot}$,
where $M_{WD}$ is the mass of the white dwarf.

\section{Discussion and Conclusions}
\label{sec:disc}



The nature of V842~Cen remains unclear. Optical data obtained in 2008 
 show modulations at 56.825 s and 3.780$\pm$0.004
h \citep{woudt}. These authors attributed the shorter period to the white dwarf
spin. The presence of sidebands at 56.598 and 57.054 s suggested that
the orbital period should be at 3.94 h. In this picture, they then attributed the 3.780
h period to a ``negative superhump,'' a hump-shaped feature in the
light curve with period a few percent less than the orbital period of
the binary. Superhumps are believed to be associated with a tilted
accretion disk. This analysis strongly suggests that V842~Cen should
be classified as an IP. It is somewhat puzzling why earlier fast optical
photometry of V842~Cen did not show periodicity
\citep{woudt03}; presumably, flaring activity observed in that data
masked any periodic  signatures. Our data cannot confirm the short
period at X-ray or optical wavelengths. 
While the lack of a $\sim$57~s period in the X-ray data does not necessarily
rule out a rapidly rotating white dwarf, one generally expects the
X-ray data from IPs to be modulated at the spin period, since the localized
accretion shock near the surface of the white dwarf produces a
high temperature thermal spectrum. Unfortunately, our
optical data are not sensitive enough to detect a pulse amplitude of
4 mmag as previously measured. 

On the other hand, we detect significant periods at 3.51$\pm$0.40 h in X-rays
and 4.0$\pm$0.8 h in optical. The pulsed fraction $p$ of the X-ray light curves folded at this period
decreases with energy, ranging from 47$\pm$13\% in the soft (0.3-1.0
keV), to 45$\pm$12\% in the medium (1.0-2.0 keV), and to 30$\pm$11\% in the
hard (2.0-10.0 keV) X-ray band. A similar trend is observed in the X-ray
spin-phased light curves of IPs \citep[e.g.,][EX~Hya]{alan} and can be caused by the energy dependence of the absorption cross-section,
occultation of the lower (cooler) portion of the accretion column by
the white dwarf's body or both. 
%
The modulation detected in our data is somehow different from
the  orbital period claimed by \citet{woudt} at the 3$\sigma$ level. 
Furthermore, the smooth shape of the phase-folded light curves indicates that an orbital origin of the
modulation is unlikely. A low orbital inclination as suggested by \citet{woudt03} also supports this conclusion.
Eclipses of the accretion column are observed
as sharp decreases in the light curves of IPs with high orbital inclination \citep[][]{ronnie},
lasting for tenths or less of the orbital period. No conspicuous eclipse feature could be found using a 50 s binning in any of the light curves.



On the other hand, the period we have detected is consistent with the
value reported by \citet{woudt} as a negative superhump; however, it
is not clear how the tilted accretion disk would contribute to such
high energy X-ray emission.  To our knowledge, negative superhumps
have not been detected in a CV at X-ray wavelengths. Furthermore, the
energy dependence of the pulsed fraction of an X-ray superhump would
require explanation.

The X-ray spectrum is fully 
consistent
with thermal emission originating in the post-shock region of a magnetic
accretion column \citep[e.g.][]{mukai03}.
The model fit to the X-ray spectrum
suggests that the white dwarf in V842~Cen has a mass $M_{WD}$=0.88
$M_{\odot}$, which is not unusual for an IP. In fact,
\cite{brunschweiger} found that among the IPs detected with
$Swift$/BAT, 5 out of 17 have white dwarfs more massive than 0.88 $M_{\odot}$.
Given the good agreement with the accretion
shock model and a reasonable mass determination, it is difficult to
reconcile the lack of detection of the fast period associated with the
white dwarf spin. Either the fast
period is undetectable due to a low inclination or
the period we detect is
instead the white dwarf spin period. The latest possibility would be, if confirmed, very rare since the detected period is longer than any known spin periods from IPs,\footnote{See the latest catalog of IPs at http://asd.gsfc.nasa.gov/Koji.Mukai/iphome/catalog/alpha.html}.



The shape of the nova remnant in V842~Cen also remains puzzling. The remnant geometry depends on the
matter distribution and the ejecta illumination by the central
source. If no accretion disk is present in V842~Cen (e.g. it
is a polar CV instead of an IP), then the  ejecta might be more
homogeneously illuminated; however,  at the moment we do not think that this is a possibility as no evidence has been reported elsewhere on the detection of optical polarization or strong soft X-ray emission that would support the polar nature.
One expects the presence of a large accretion disk to affect the photoionization
in different regions of the nebula. The accretion disk may absorb the
ionizing photons in the equatorial regions causing an aspherical
illumination. In HR~Del for example, the observed aspherical
illumination is attributed to high mass transfer and a large
disk  \citep{moraes09}.


Furthermore, a high white dwarf rotational velocity can generate asymmetries in
the eruption conditions because of the effective gravity variation
from equatorial to polar regions \citep{scott00}. If we consider a
0.88 M$_{\odot}$ white dwarf with a rotational period of 57 s, we should obtain
ejecta with axial ratio of $\sim$1.35 \citep[using expressions 8 to
11 by][]{scott00}. If the rotational period is 3.51 h, the envelope
would be symmetric. A period P$\lesssim$125 s is required to detect an
asymmetric envelope. If the white dwarf is more massive than what we
have derived, the asymmetry would be
smaller, but detectable (as a prolate remnant). 


In summary, the X-ray spectrum and the energy dependence of its pulse
fraction support the classification of V842~Cen as an IP, but the
symmetric shape of the ejecta and the long period of the light curves are not easily explained in this picture. 
While the X-ray
data do not
definitively rule out a very fast 57~s spin period, a longer period
would be less discrepant with the ejecta symmetry.
Another possibility is that the ejecta
asymmetry axis is the same as that of the line of  sight. 
A clear, definitive
determination of the nature of V842~Cen, in particular whether or not
it is an IP and what its spin period is, will contribute to
understanding the system and its ejecta morphology.

\section*{Acknowledgments}

We acknowledge the anonymous referee for the comments that helped to improve the manuscript. We acknowledge Raimundo Lopes de Oliveira for useful tips and discussion about data analysis. Based on observations obtained with XMM-Newton, an ESA science mission with instruments and contributions directly funded by ESA member states and NASA.

\bsp

\label{lastpage}

\end{document}